\documentstyle[pre,preprint,aps]{revtex}
\begin{document}
\draft
\title{Tunneling and the Onset of Chaos in a Driven Bistable System}
\author{R. Utermann, T. Dittrich, and P. H\"anggi}
\address{Institut f\"ur Physik, Universit\"at Augsburg, Memminger Stra\ss e 6,
              D-86135 Augsburg, Germany}
\date{\today}
\maketitle
\begin{abstract}
We study the interplay between coherent transport by tunneling and diffusive
transport through classically chaotic phase-space regions, as it is reflected
in the Floquet spectrum of the periodically driven quartic double well. The
tunnel splittings in the semiclassical regime are determined with high
numerical accuracy, and the association of the corresponding doublet states
to either chaotic or regular regions of the classical phase space is
quantified in terms of the overlap of the Husimi distribution with the
chaotic layer along the separatrix. We find a strong correlation between
both quantities. They show an increase by orders of magnitude as chaotic
diffusion between the wells starts to dominate the classical dynamics. We
discuss semiclassical explanations for this correlation.
\end{abstract}
\pacs{05.45.+b, 03.65.Sq, 73.40.Gk}

\narrowtext
\section{Introduction}
\label{s1}
In mesoscopic systems with a bistable potential, two fundamentally different
modes of transport can occur simultaneously: Coherent quantal transport by
tunneling, and deterministic diffusion through chaotic regions interweaving the
classical phase space. There is a strong mutual influence of these two
processes. While the classical phase-space structure is clearly reflected in
the quantal spectrum \cite{Z,BG,B}, localization and delocalization due to
quantal interference, in turn, effectively alter the boundary conditions for
the quantal counterpart of classical phase-space diffusion
\cite{GCIF,FGP,She,DH,BTU1,TU,BTU2}.
\par
The idea that chaotic layers separating symmetry-related regular regions can
give rise to quantal tunneling, just as potential barriers do, has been
introduced by Davis and Heller \cite{DH}, who dubbed this notion ``dynamical
tunneling''. The fingerprint of dynamical tunneling in the spectrum has been
studied by Bohigas {\it et al.\/} \cite{BTU1,TU}. They utilized the occurrence
of tunnel doublets with exponentially small splittings as a filter to separate
``regular'' quantal eigenstates (i.e., those localized in classically regular
phase-space regions) from ``chaotic'' ones (in an analogous sense).  While
their work was based on a time-independent, two-dimensional nonlinear
oscillator, the first inquiry into driven tunneling in a one-dimensional,
bistable system was undertaken by Lin and Ballentine \cite{LB} and Grossmann
{\it et al.\/} \cite{GJDH1,GJDH2,GDH}, and subsequently by Plata and Gomez
Llorente \cite{PG}.
\par
Up to now, the only attempt towards an analytical understanding of tunneling
between quantum states localized on symmetry-related tori (or chaotic regions)
in multi-dimensional systems has been made by Wilkinson and Hannay \cite{W,WH}.
Using semiclassical methods, they express the tunnel splitting as a function of
the classical action along paths connecting the classical manifolds that
support the doublet states. However, their theory does not provide a unified
description spanning the transition from regular to chaotic states. It is this
transition which particularly interests us here.
\par
Our present work (a preliminary account of which has been published in ref.
\cite{UDH}) is similar in scope to that of Bohigas {\it et al.\/} mentioned
above, but devoted to a different system class.  Applying torus or EBK
(Einstein-Brillouin-Keller) quantization for periodically driven systems
\cite{BH} as well as the random-matrix theory for mixed (regular and chaotic)
systems \cite{BTU1,BR} to the Floquet spectrum of the periodically driven
double well, we expect the following scenario: The classical phase space of
this system, for not too strong driving, is characterized by a pair of
symmetry-related regular regions near the minima of the unperturbed potential,
embedded in a chaotic layer along the unperturbed separatrix.  Quantal
eigenstates localized in the regular regions form tunnel doublets with
exponentially small splittings.  These doublets will break up, and their
splittings will reach values of the order of the mean level spacing, as soon as
the associated pair of quantizing tori dissolves in the chaotic layer.
Therefore, the tunnel splitting should depend strongly on the nature---regular
or chaotic---of the classical phase-space region to which the corresponding
pair of eigenstates predominantly belongs. In order to allow for a numerical
test of this conclusion, we shall quantify this association in terms of the
overlap of a suitable phase-space representation of the eigenstates with the
chaotic layer.
\par
We emphasize that our hypothesis relies crucially on the specific phase-space
topology of driven bistable systems. That is, the decisive property that
prevents states within the chaotic layer from forming tunnel doublets is not a
positive local Lyapunov exponent of the classical dynamics, but rather the
possibility of phase-space transport across the symmetry plane. For example, in
the converse case---symmetry-related chaotic islands embedded in regular
regions---one expects tunnel-splitted level pairs which, in turn, repel each
other in the same way as the single levels of the individual chaotic regions
would do. Furthermore, it should be kept in mind that the disintegration of a
classical torus is not an abrupt all-or-nothing event.  Rather, this invariant
manifold passes through a fractal, intermediate stage (``cantorus'' \cite{Rl})
before it disappears, and even after that, a ``shadow'' remains in the form of
a distinct, repelling structure within the chaotic sea (``vague torus''
\cite{RS}). Due to the finite phase-space resolution of the quantal dynamics,
this transition is smeared out even further in the parameter space of the
corresponding quantum system.
\par
We give a brief review of our working model, its symmetries and classical
dynamics, in Section \ref{s2}. In Section \ref{s3}, we introduce the methods
and key quantities used in our numerical study and interpret our results on
basis of the simple picture just sketched. Section \ref{s4} contains the
summary and puts our findings in a broader perspective.

\section{The Model: Symmetries and Classical Dynamics}
\label{s2}
The harmonically driven quartic double well is described by the Hamiltonian
\begin{eqnarray}
H(x,p;t) &=& H_0(x,p) + H_1(x;t),\nonumber\\
H_0(x,p) &=& {p^2 \over 2} - {1 \over 4} x^2 +{1 \over 64 D} x^4, \\
H_1(x;t) &=& x\,S\,\cos (\omega t).\nonumber
\end{eqnarray}
With the dimensionless variables used, the only parameter controlling the
unperturbed Hamiltonian $H_0(x,p)$ is the barrier height $D$. It can also be
interpreted as the (approximate) number of doublets with energies below the top
of the barrier. Accordingly, the classical limit amounts to letting $D
\rightarrow \infty$. The driving is characterized by its amplitude $S$ and
frequency $\omega$. In all our numerical studies, we kept barrier height and
driving frequency fixed at the values $D = 8$ and $\omega = 0.95$,
respectively.
\par
Besides invariance under time-reversal, $x \rightarrow x$, $p \rightarrow
-p$, $t \rightarrow -t$, the unperturbed system possesses the spatial
reflection symmetry $x \rightarrow -x$, $p \rightarrow p$, $t \rightarrow t$.
For a general periodic driving, this twofold symmetry is destroyed and
replaced by the discrete time-translation invariance under $t \rightarrow t
+ 2\pi / \omega$. On the quantum-mechanical level, this enables the
application of the Floquet formalism \cite{Shi,Sa,MOR,Ch}. For the specific
time dependence of a harmonic driving, the symmetry $f(t + \pi / \omega) =
-f(t)$ restores a similar situation as in the unperturbed case: The system is
now invariant against the operation \cite{GJDH1,GJDH2,GDH,P}
\begin{equation}
\mbox{\sf P}: p \rightarrow -p, \quad x \rightarrow -x, \quad
t \rightarrow t + {\pi \over \omega},
\end{equation}
which may be regarded as a generalized parity in the extended phase space
spanned by $x$, $p$, and phase, i.e., time $t \bmod (2\pi / \omega)$.  As in
the unperturbed case, this enables to separate the eigenstates into an even and
an odd subset. Moreover, as far as the dynamics allows, the generalized parity
gives rise to tunnel doublets with the corresponding pair of eigenstates
residing within the vortex tubes \cite{BH} formed by the motion within the
potential wells of the unperturbed system. These tunnel doublets, in
particular, are the subject of our study.
\par
The unperturbed system, as given by $H_0(x,p)$, exhibits the generic phase
space of a bistable system (Fig.\ 1a). Its structure is even independent, up to
a rescaling of $x$ and $p$, of the parameter $D$, provided $D > 0$. Apart from
a constant of the order of unity, $D$ gives the phase-space area enclosed by
the two loops of the separatrix. It represents the characteristic action of the
unperturbed system to be compared, in particular, with $\hbar$.
\par
For the parameter values of the driving that are relevant for the present
study, $\omega = 0.95$ and $0 \leq S \leq 0.2$, the phase space of the
unperturbed system is modified mainly by two additional features \cite{RZ}: The
onset of chaos in the vicinity of the separatrix, and the growth of the first
nonlinear resonance zone (Figs.\ 1b,c). The chaotic layer along the separatrix
develops out of the homoclinic tangle, the intricate interweaving of the stable
and the unstable manifolds originating at the hyperbolic fixed point near the
top of the barrier \cite{LL}. It is present for all $S > 0$, but covers a
substantial part of phase space only for $S \gtrsim 0.1$.
\par
Nonlinear resonance arises when (an integer multiple $k$ of) the period of the
driving equals the period of the oscillation within each well at some energy.
Each resonance generates a chain of $k$ alternating elliptic and hyperbolic
fixed points, embedded in the regular regions within each well. For example,
the vortex tube surrounding, in the extended phase space, the elliptic fixed
point of the first resonance, winds around the vortex tube corresponding to
nonresonant motion once per period of the driving. Within the parameter
interval $10^{-3} \lesssim S \lesssim 10^{-2}$, this first resonance zone grows
from near invisibility to almost the full area of the regular region within
each well. The higher resonances occupy only extremely narrow filaments in
phase space which accumulate at the top of the barrier. Therefore, the border
zone between the chaotic layer and the regular regions, formed by the higher
resonances, is also very narrow, and in fact comes close to a sharply defined
borderline \cite{BDH}.  As a further consequence, the chaotic layer ``gains
ground'' in a quite smooth manner, as $S$ increases. The intricate,
self-similar layering of the border zone implied by the KAM theorem \cite{LL}
does not visibly affect this process.
\par
For larger values of the driving amplitude, higher-order features, such as
secondary islands, appear and render the phase-space structure increasingly
complicated (Fig.\ 1d). As we shall see below, the
growth of the first resonance
leaves no significant trace in the tunnel splittings, while the consequences of
the spreading of the chaotic layer are drastic.

\section{Numerical Methods and Results}
\label{s3}
The basic ingredient required to study the quantum-mechanical aspects of a
periodically driven system is the Floquet operator \cite{Shi,Sa,MOR,Ch}, i.e.,
the unitary propagator that generates the time evolution over one period of the
driving force,
\begin{equation}
U = \mbox{\sf T} \exp \left( -{\mbox{\rm i} \over \hbar}
\int_0^{2\pi / \omega} \mbox{\rm d} t \, H(t) \right),
\end{equation}
where $\mbox{\sf T}$ denotes time ordering. Its eigenvectors and eigenphases,
referred to as Floquet states and quasienergies, respectively, can be written
in the form
\begin{equation}
|\,\psi_{\alpha}(t)\,\rangle = \mbox{\rm e}^{-\mbox{\scriptsize\rm i}
\epsilon_{\alpha}t} |\,\phi_{\alpha}(t)\,\rangle,
\end{equation}
with
\begin{eqnarray*}
|\,\phi_{\alpha}(t + 2\pi / \omega)\,\rangle = |\,\phi_{\alpha}(t)\,\rangle.
\end{eqnarray*}
{}From a Fourier expansion of the $|\,\phi_{\alpha}(t)\,\rangle$,
\begin{eqnarray}
\begin{array} {rcl}
{\displaystyle |\,\phi_{\alpha}(t)\,\rangle} &=&
{\displaystyle \sum_n |\,c_{\alpha,n}\,\rangle
\mbox{\rm e}^{-\mbox{\scriptsize\rm i} n\omega t},} \\
{\displaystyle |\,c_{\alpha,n}\,\rangle} &=&
{\displaystyle {\omega \over 2\pi} \int_0^{2\pi / \omega} \mbox{\rm d} t \,
|\,\phi_{\alpha}(t)\,\rangle \mbox{\rm e}^{\mbox{\scriptsize\rm i} n\omega t},}
\end{array}
\end{eqnarray}
it is obvious that the quasienergies come in classes, $\epsilon_{\alpha,n} =
\epsilon_{\alpha} + n\omega$, $n = 0, \pm 1, \pm 2, \ldots$, where each member
corresponds to a physically equivalent solution. Therefore, all spectral
information is contained in a single ``Brillouin zone'', $-\omega/2 \leq
\epsilon < \omega/2$.
\par
By inserting the eigenstates (4) into the Schr\"odinger equation, Fourier
expanding, and using the representation in the eigenstates of the unperturbed
Hamiltonian, $H_0 |\,\Psi_k\,\rangle = E_k |\,\Psi_k\,\rangle$, a matrix
eigenvalue equation \cite{Shi,Ch},
\begin{equation}
\sum_{n'} \sum_{k'} H_{n,k;n',k'} c_{n',k'} = \epsilon c_{n,k}
\end{equation}
is derived, where
\begin{eqnarray*}
H_{n,k;n',k'} &=& (E_k - n\omega) \delta_{n-n'} \delta_{k-k'} \\
&&+ S \, x_{k,k'} \, {\delta_{n-1-n'} + \delta_{n+1-n'} \over 2}, \\
c_{n,k} &=& \langle\,\Psi_k\,|\,c_n\,\rangle, \\
x_{k,k'} &=& \langle\,\Psi_{k'}\,|\,x\,|\,\Psi_k\,\rangle.
\end{eqnarray*}
It is this matrix eigenvalue equation which we solve numerically to obtain
the Floquet states and the quasienergies.
\par
The invariance of the system under the generalized parity discussed in the
previous section (see Eq.\ (2)) is of considerable help in the treatment of
Eq.\ (6). The eigenvalue equations for the subspaces spanned by the even and
odd eigenvectors decouple completely, and the matrices to be diagonalized take
the block structure
\widetext
\begin{mathletters}\label{eq:all}
\begin{equation}
H_{\mbox{\rm\scriptsize e}} =  \left( \begin{array} {ccccccc}
&\vdots&\vdots&\vdots&\vdots&\vdots&               \\
\cdots                                   &E_{\mbox{\rm\scriptsize e}}+
2\omega I&X_{\mbox{\rm\scriptsize eo}}&0&0&0&\cdots\\
      \cdots&X_{\mbox{\rm\scriptsize eo}}&E_{\mbox{\rm\scriptsize o}}+
   \omega I&X_{\mbox{\rm\scriptsize oe}}&0&0&\cdots\\
    \cdots&0&X_{\mbox{\rm\scriptsize oe}}&E_{\mbox{\rm\scriptsize e}}
             &X_{\mbox{\rm\scriptsize eo}}&0&\cdots\\
  \cdots&0&0&X_{\mbox{\rm\scriptsize eo}}&E_{\mbox{\rm\scriptsize o}}-
       \omega I&X_{\mbox{\rm\scriptsize oe}}&\cdots\\
\cdots&0&0&0&X_{\mbox{\rm\scriptsize oe}}&E_{\mbox{\rm\scriptsize e}}-
        2\omega I                           &\cdots\\
&\vdots&\vdots&\vdots&\vdots&\vdots&
\end{array} \right), \label{eq:a}
\end{equation}
\begin{equation}
H_{\mbox{\rm\scriptsize o}} = \left( \begin{array} {ccccccc}
&\vdots&\vdots&\vdots&\vdots&\vdots&               \\
\cdots                                   &E_{\mbox{\rm\scriptsize o}}+
2\omega I&X_{\mbox{\rm\scriptsize oe}}&0&0&0&\cdots\\
      \cdots&X_{\mbox{\rm\scriptsize oe}}&E_{\mbox{\rm\scriptsize e}}+
   \omega I&X_{\mbox{\rm\scriptsize eo}}&0&0&\cdots\\
    \cdots&0&X_{\mbox{\rm\scriptsize eo}}&E_{\mbox{\rm\scriptsize o}}
             &X_{\mbox{\rm\scriptsize oe}}&0&\cdots\\
  \cdots&0&0&X_{\mbox{\rm\scriptsize oe}}&E_{\mbox{\rm\scriptsize e}}-
       \omega I&X_{\mbox{\rm\scriptsize eo}}&\cdots\\
\cdots&0&0&0&X_{\mbox{\rm\scriptsize eo}}&E_{\mbox{\rm\scriptsize o}}-
        2\omega I                           &\cdots\\
&\vdots&\vdots&\vdots&\vdots&\vdots&
\end{array} \right), \label{eq:b}
\end{equation}
\end{mathletters}
for the even ($\mbox{\rm e}$) and the odd ($\mbox{\rm o}$) subspace,
respectively, where
\begin{eqnarray*}
E_{\mbox{\rm\scriptsize e}} = \left( \begin{array} {cccc}
                 E_0     &0     &0     &\cdots\\
                   0     &E_2   &0     &\cdots\\
                   0     &0     &E_4   &\cdots\\
                   \vdots&\vdots&\vdots&\ddots
\end{array} \right), \quad
E_{\mbox{\rm\scriptsize o}} = \left( \begin{array} {cccc}
                 E_1     &0     &0     &\cdots\\
                   0     &E_3   &0     &\cdots\\
                   0     &0     &E_5   &\cdots\\
                   \vdots&\vdots&\vdots&\ddots
\end{array} \right),
\end{eqnarray*}
\begin{eqnarray*}
X_{\mbox{\rm\scriptsize eo}} = {1 \over 2} \left( \begin{array} {cccc}
                                 x_{0,1}&x_{0,3}&x_{0,5}&\cdots\\
                                 x_{2,1}&x_{2,3}&x_{2,5}&\cdots\\
                                 x_{4,1}&x_{4,3}&x_{4,5}&\cdots\\
                                 \vdots&\vdots&\vdots&\ddots
\end{array} \right), \quad
X_{\mbox{\rm\scriptsize oe}} = {1 \over 2} \left( \begin{array} {cccc}
                                 x_{1,0}&x_{1,2}&x_{1,4}&\cdots\\
                                 x_{3,0}&x_{3,2}&x_{3,4}&\cdots\\
                                 x_{5,0}&x_{5,2}&x_{5,4}&\cdots\\
                                 \vdots&\vdots&\vdots&\ddots
\end{array} \right) ,
\end{eqnarray*}
\narrowtext
All the matrices involved are formally infinite and have to be truncated
according to the required numerical accuracy. Evidently, exploiting the
generalized parity reduces the linear dimensions of vectors and matrices by a
factor of two.
\par
The tunnel doublets we are looking for are then composed of one state each from
the even and the odd subspace, and the corresponding tunnel splittings are
given by
\begin{equation}
\Delta_l = |\epsilon_{l,\mbox{\rm\scriptsize o}} -
\epsilon_{l,\mbox{\rm\scriptsize e}}|.
\end{equation}
Due to its periodic, Brillouin-zone structure, the quasienergy spectrum gives
no hint how to order the doublets. Instead, it is possible to associate a mean
energy to the eigenstates by the relation \cite{Ch}
\begin{equation}
\bar E_{\alpha} = {\omega \over 2\pi} \int_0^{2\pi / \omega} \mbox{\rm d} t \,
\langle\,\psi_{\alpha}(t)\,|\,H(t)\,|\,\psi_{\alpha}(t)\,\rangle,
\end{equation}
which is defined on the positive real axis and thus provides a basis for an
ordering of the Floquet states. Using this order, we are able to pick out the
doublets from the ``ground-state'' one upward to that immediately below the top
of the barrier. In the following, we shall concentrate on this set of states,
which for $D = 8$ are 18 in number.
\par
A quantity which facilitates making contact with the classical dynamics is the
Husimi distribution \cite{H,G}. It is defined as the overlap of a given state
with a minimum-uncertainty (coherent) state localized at a position $(x,p)$ in
phase space,
\begin{equation}
Q_{\alpha}(x,p;t) = {1 \over 2\pi \hbar} \left|
\langle\,\xi\,|\,\psi_{\alpha}(t)\,\rangle \right|^2,
\end{equation}
with
\begin{eqnarray*}
|\,\xi\,\rangle = \mbox{\rm e}^{-|\xi|^2/2}
\sum_{n=0}^{\infty} {\xi^n \over n!} |\,n\,\rangle,
\end{eqnarray*}
where $\xi = x + \mbox{\rm i} p$, and $|\,n\,\rangle$ is an eigenstate of the
harmonic oscillator $(p^2 + x^2)/2$. The Husimi distribution represents a
quantum state as a proper phase-space probability distribution with the highest
resolution allowed by the uncertainty principle.
\par
In particular, the Husimi distribution can be used to quantify the notion of an
eigenstate residing predominantly in a specific (regular or chaotic) region of
classical phase space.  We define the mean overlap of a state
$|\,\psi_{\alpha}(t)\,\rangle$ with the chaotic layer as
\begin{equation}
\bar \Gamma_{\alpha} = {\omega \over 2\pi}
\int_0^{2\pi / \omega} \mbox{\rm d} t
\int_{-\infty}^{\infty} \mbox{\rm d} x
\int_{-\infty}^{\infty} \mbox{\rm d} p \,
Q_{\alpha}(x,p;t) \Gamma(x,p;t).
\end{equation}
Here, $\Gamma(x,p;t)$ denotes the characteristic function for the chaotic
region in the vicinity of the separatrix. Since the Husimi distribution forms a
normalized probability distribution over phase space, we have $0 \leq \bar
\Gamma_{\alpha} \leq 1$. The characteristic function can be determined
numerically, e.g., by letting a trajectory started anywhere in this chaotic
region ``tick'' boxes in a coarse-grained phase space of the desired
resolution.
\par
We start the discussion of our numerical results with a phenomenology of the
Floquet states involved in driven tunneling. There are three aspects to be
mentioned, the dependence on the quantum number (we choose the doublets 1 to
7), on the amplitude $S$ of the driving, and on time $t\bmod(2\pi / \omega)$.
For small enough $S$, e.g., $S = 10^{-5}$, the Floquet states are nearly
identical with the unperturbed eigenstates (Fig.\ 2). They are localized on
tori with increasing characteristic action, embedded in the regular regions in
the wells. The first qualitative changes to occur with $S$ taking larger values
are due to the growth of the first resonance (see the previous section).  For
$S = 10^{-2}$ (Fig.\ 3), the doublet states with $l = 2,4,5$ take shapes that
indicate their localization on tori that belong to the first resonance rather
than to nonresonant motion in the wells; the characteristics of torus
quantization remain. At $S = 0.2$, the chaotic layer occupies an appreciable
part of phase space. For the higher-lying doublets, the distorted toroidal
shape of the Husimi distribution then gives way to a more rugged form which
less closely resembles a smeared-out torus (Fig.\ 4). A glance at the time
dependence within one period complements these observations (Fig.\ 5): For a
state pertaining to the first resonance, e.g., the two maxima of the Husimi
distribution rotate clockwise, with the frequency of the driving, around the
respective potential minima, as do the corresponding classical vortex tubes.
\par

The central result of our study is presented in Fig.\ 6, where we compare the
$S$
dependence of the tunnel splittings $\Delta_l$ (part a) with that of the
overlaps $\bar \Gamma_{l,\mbox{\rm e}}$ (part b; we could quite as well have
chosen $\bar \Gamma_{l,\mbox{\rm o}}$). Looking at the tunnel splittings first,
a relatively clear-cut picture emerges: For $S \lesssim 10^{-3}$, the tunnel
splittings do not deviate significantly from their unperturbed values (apart
from an irregularity in $\Delta_2$ which roughly coincides with a region of
rapid growth of the first resonance). They increase roughly exponentially with
the quantum number $l$, from a value $\Delta_1 = 5.20 \times 10^{-18}$, in good
agreement with the corresponding semiclassical (instanton-method, cf.
ref.\cite{Co}) estimate \cite{GJDH1} $\Delta_1^{\mbox{\scriptsize\rm sc}} =
\sqrt{128D/\pi}\,\exp(-16D/3) = 5.33 \times 10^{-18}$, to $\Delta_8 = 5.59
\times 10^{-4}$. For $S \gtrsim 10^{-3}$, the splittings start one by one to
grow exponentially with $S$, from the ground-state doublet ($l = 1$) upwards,
so that at $S = 0.2$, the range of the splittings has shrunk from 14 to three
orders of magnitude. The $S$ dependence of the overlaps shows the same
qualitative features as that of the tunnel splittings: There is no significant
deviation from the unperturbed values for $S \lesssim 10^{-3}$, while for
larger $S$, exponential growth sets in from bottom to top of the doublet
ladder, so that the $\bar \Gamma_{l,\mbox{\rm e}}$ successively join in an
approximately single line.
\par
The qualitative agreement between the two respective groups of functional
dependences indicates that {\it there exists a strong correlation between the
tunnel splittings and the overlaps with the chaotic layer\/}. Furthermore, the
steep exponential increase occurring in both quantities coincides with the
onset of chaotic motion in the classical dynamics, whereas, e.g., the transfer
of phase-space area from nonresonant motion to the first resonance hardly
leaves any trace in the tunnel splittings. Insofar, the simple picture sketched
in the Introduction is confirmed. Details of our expectation, however, need to
be revised.
\par
In particular, the notion that each splitting widens up individually as the
corresponding quantizing torus resolves, is not unambiguously corroborated by
the data. It would imply that the transitions to a large splitting occur from
top to bottom, i.e., first for the doublet localized on the outermost torus,
the one with the highest mean energy. Indeed, if this transition is assessed
from the splittings passing a certain absolute threshold, say $\Delta_l =
10^{-4}$, that order is roughly followed (at least for those doublets we can
keep track of that far). If, however, the point of onset of exponential growth,
visible in a logarithmic plot, is taken as the criterion, the order is
reversed.
\par
Another remarkable fact is that the widening of the splittings, and the
concomitant change in character of the eigenfunctions, as a function of $S$, is
a continuous process that can only vaguely be associated with the decay of a
KAM torus, taken as as a discrete event. Even doublet states overlapping by
$70\%$ with the chaotic layer may still show a relatively small splitting and
exhibit the signature of a regular state in their spatial structure and time
dependence (see, e.g., Figs.\ 4 and 7). It remains to be clarified whether this
retarded decay of the tunnel doublets corresponds to the gradual
disintegration of classical tori via cantori and vague tori.
\par
Unfortunately, our numerical means do not allow to extend the data beyond $S =
0.2$. Therefore, the eventual saturation of the growth of the tunnel
splittings, when they reach the order of magnitude of the mean level spacing,
could not be studied. Furthermore, even if the double well with a scaled
barrier height $D = 8$ is the closest approach to the classical limit we could
practically afford, the quantal dynamics of this system is not yet sensitive
enough for the details of the classical phase space to allow for a conclusive
numerical test of a semiclassical description.

\section{Summary}
\label{s4}
In which way does the onset of classical chaos in the vicinity of the
separatrix influence the tunnel splittings in a periodically driven bistable
system? The random-matrix theory for mixed (chaotic and regular) systems,
together with semiclassical considerations, suggest the following simple
answer: A tunnel splitting undergoes a transition from exponentially small
values to a size comparable with the mean level spacing, as the corresponding
pair of symmetry-related, quantizing tori resolve in a common chaotic region.
In order to test this hypothesis, we quantified the association of an
eigenstate to the chaotic layer along the separatrix in terms of the overlap of
its Husimi representation with this phase-space region. We found a striking
qualitative agreement between the functional dependences of the splittings and
the corresponding overlaps on the amplitude of the driving force. Both groups
of quantities show transitions to steep exponential growth as the chaotic layer
along the separatrix spreads at the expense of the regular regions around the
potential wells. The above hypothesis does, however, imply more, namely that
the transitions to a large splitting for the individual doublets occur in a
specific order given by the fate of the corresponding tori.  For the driven
double well this order is roughly from top to bottom on the energy axis (where
``energy'' should be read as ``mean energy''). This is indeed what we observe
if the widening of the splittings is assessed from their growing beyond some
absolute threshold value. If, however, the onset of exponential growth is taken
as the criterion, the order turns out to be from bottom to top. As a further
unexpected detail, the regular character of doublet states survives until far
into a parameter region where they are already located amidst the chaotic
layer.
\par
On first sight it may appear strange that a static quantity like the overlap
with the chaotic layer should bear on a transport phenomenon such as tunneling.
{}From a semiclassical point of view, this may become understandable if one
assumes that a path-integral expression for the tunnel splitting becomes
dominated by contributions from paths which pass through the chaotic layer, as
soon as chaotic diffusion enables significant classical transport between the
wells. Our study, which is intended solely as a first numerical survey, should
encourage to cast this notion in a precise semiclassical form. Related topics
which deserve further investigation are the time-domain aspects of tunneling
through the chaotic layer, the occurrence of crossings between tunnel doublets
and their influence on the dynamics. Another interesting open question is
whether states which reside within the chaotic layer but appear regular in
character, can be associated with vague tori of the classical dynamics.



\begin{figure}
\caption{
Classical phase-space portraits of the periodically driven double well at phase
0 of the driving, for various values of the driving amplitude. The parameter
values are $D = 8$, $\omega = 0.95$, and (a) $S = 0$, (b) $S = 10^{-3}$, (c) $S
= 0.02$, (d) $S = 0.2$.}
\label{f1}
\end{figure}

\begin{figure}
\caption{
Contour plots of the Husimi distributions for the Floquet states
$|\,\psi_{2,\mbox{\rm\scriptsize e}}(0)\,\rangle$ (a) and $|\,\psi_{7,
\mbox{\rm\scriptsize
e}}(0)\,\rangle$ (b), at $S = 10^{-5}$.}
\label{f2}
\end{figure}

\begin{figure}
\caption{
Contour plot of the Husimi distribution for the Floquet state
$|\,\psi_{5,\mbox{\rm\scriptsize e}}(0)\,\rangle$ at $S = 0.02$. It should be
compared
with the corresponding classical phase-space portrait, Fig.\ 1c.}
\label{f3}
\end{figure}

\begin{figure}
\caption{
Contour plot of the Husimi distribution for the Floquet state
$|\,\psi_{7,\mbox{\rm\scriptsize e}}(0)\,\rangle$ at $S = 0.2$.}
\label{f4}
\end{figure}

\begin{figure}
\caption{
Contour plot of the Husimi distribution for the Floquet state
$|\,\psi_{5,\mbox{\rm\scriptsize e}}(t)\,\rangle$ at $S = 0.02$ and
(a) $\omega t = \pi/4$
and (b) $\omega t = \pi/2$ (b), compared to the corresponding classical
phase-space portraits at the same parameter values and phases (parts (c) and
(d), respectively). For the initial states ($\omega t = 0$), see Figs.\ 1c and
3, respectively. The states at later times $\omega t = n \pi/4$ are related to
those shown by simple phase-space symmetries, as implied by the generalized
parity, Eq.\ (2).}
\label{f5}
\end{figure}

\begin{figure}
\caption{
Tunnel splittings (a) and overlaps with the chaotic layer (b) for the seven
lowest tunnel doublets, as functions of the amplitude of the driving.}
\label{f6}
\end{figure}

\begin{figure}
\caption{
Contour plot of the Husimi distribution for the Floquet state
$|\,\psi_{7,\mbox{\rm\scriptsize e}}(t)\,\rangle$ at $S = 0.02$ and
(a) $\omega t = \pi/4$
and (b) $\omega t =
\pi/2$. For the initial state, see Fig.\ 4. The states at later times
$\omega t = n \pi/4$ are related by
simple phase-space symmetries to those shown here, as implied by the
generalized parity,Eq.\ (2).}
\label{f7}
\end{figure}

\end{document}